\documentclass[aip,jap,showpacs,showkeys,nofootinbib,floatfix,reprint]{revtex4-1}
\usepackage{graphicx}	
\usepackage{amsmath}
\usepackage{subfigure}
\usepackage{units}
\usepackage{braket}

\begin{document}

\newcommand{\dd}{\textrm{d}}

\title{Optical absorption in highly-strained Ge/SiGe quantum wells: the role of $\Gamma\rightarrow\Delta$ scattering}

\author{L.~Lever}
\email{l.j.m.lever@leeds.ac.uk}
\author{Z.~Ikoni\'{c}}
\author{A.~Valavanis}
\author{R.~W.~Kelsall}
\affiliation{Institute of Microwaves and Photonics, School of Electronic
and Electrical Engineering, University of Leeds, Leeds LS2 9JT, United
Kingdom}
\author{M.~Myronov}
\author{D.~R.~Leadley}\affiliation{Department of Physics, University of Warwick, Coventry CV4 7AL, United Kingdom}
\author{Y.~Hu}
\author{N.~Owens}
\author{F.~Y.~Gardes}
\author{G.~T.~Reed}
\affiliation{Optoelectronics Research Centre,
University of Southampton,
Highfield Campus,
Southampton,
SO17 1BJ,
UK}

\begin{abstract}
We report the observation of the quantum-confined Stark effect in Ge/SiGe multiple quantum well heterostructures grown on Si$_{0.22}$Ge$_{0.78}$ virtual substrates.  The large compressive strain in the Ge quantum well layers caused by the lattice mismatch with the virtual substrate results in a blue shift of the direct absorption edge, as well as a reduction in the $\Gamma$-valley scattering lifetime because of strain-induced splittings of the conduction band valleys.  
We investigate theoretically the $\Gamma$-valley carrier lifetimes by evaluating the $\Gamma\rightarrow L$ and $\Gamma\rightarrow \Delta$ scattering rates in strained Ge/SiGe semiconductor heterostructures.  These scattering rates are used to determine the lifetime broadening of excitonic peaks and the indirect absorption in simulated absorption spectra, which are compared with measured absorption spectra for quantum well structures with systematically-varied dimensions.  We find that $\Gamma\rightarrow \Delta$ scattering is significant in compressively strained Ge quantum wells and that the $\Gamma$-valley electron lifetime is less than 50\,fs in the highly-strained structures reported here, where $\Gamma\rightarrow \Delta$ scattering accounted for approximately half of the total scattering rate.
\end{abstract}

\maketitle

\section{Introduction}
\label{sec:intro}
The optical properties of direct band gap semiconductor quantum wells have been studied extensively over the past few decades, and semiconductor quantum well heterostructures are found in a number of commercial optoelectronic devices, including lasers and optical modulators.  Recently, the optoelectronic properties of Ge/SiGe quantum well structures grown on silicon substrates have become the subject of growing research interest.  The quantum-confined Stark effect (QCSE) was first reported in this system in 2005,\cite{Kuo2005} where a multiple quantum well (MQW) structure was grown on an alloy \emph{virtual substrate}, within  which strain relaxes via the insertion of misfit dislocations and where the virtual substrate composition was chosen to provide strain balance to the MQW layers so that no net strain accumulates in the MQW stack.  In the following years the QCSE has been reported using a variety of quantum well dimensions and virtual substrate compositions,\cite{Kuo2006,Schaevitz2008,Lange2009,Lever2011} and has been extended to configurations for modulator structures,\cite{Roth2008,Ren2011,Edwards2011} as well as to investigations of the carrier dynamics\cite{Lange2009,Claussen2010} and to photoluminescence studies.\cite{Gatti2011,Giorgioni2011}

In III-V quantum well systems the $\Gamma$-valley electron lifetimes are relatively long ($>$\,1\,ps), and consequently the exciton lineshapes can be described by a Gaussian profile.\cite{Chelma1984}  In Ge/SiGe quantum wells however, $\Gamma$-valley electrons are able to scatter into the lower-lying indirect valleys, and this results in a reduction of the electron lifetimes.  
The scattering lifetime is significant, as it determines the strength of the indirect absorption as well as the amount of lifetime broadening of the excitonic peaks.  
Two of the most important metrics of an optical modulator are the insertion loss and extinction ratio that can be achieved.\cite{Wakita1997}  The scattering lifetime becomes a fundamental limiting factor for the insertion loss of a Ge/SiGe quantum well modulator because these broadening mechanisms introduce losses in the structure at the applied field which corresponds to the transmitting state of the modulator.  This in turn affects the extinction ratio, since in the design of a practical device there is a trade-off to between the extinction ratio and the insertion loss that can be tolerated.

Typically, contrast in the absorption coefficient that has been achieved with Ge/SiGe quantum wells has been at wavelengths around 1450\,nm;\cite{Kuo2006,Roth2007,Schaevitz2008, Chaisakul2010} however, most telecommunications applications use wavelengths close to 1550\,nm or 1310\,nm.  Some progress has been made in getting absorption coefficient contrast at 1550\,nm,\cite{Kuo2006,Roth2008} but this has required elevated temperatures to exploit the temperature-induced bandwidth narrowing.  Recently, operation at 1310\,nm has been reported by exploiting compressive in-plane strain to induce a blue-shift of the direct band edge;\cite{Lever2010,Rouifed2012} however, it remains unclear what the impact of the strain is on the carrier dynamics, and hence on the indirect absorption and lifetime broadening of excitonic peaks.  
In this work, strain-balanced Ge/Si$_{0.4}$Ge$_{0.6}$ multiple quantum well stacks were epitaxially grown on Si$_{0.22}$Ge$_{0.78}$ virtual substrates on (100) silicon wafers in order to investigate the absorption properties of Ge quantum wells with large amounts of compressive in-plane strain. 
The measured absorption spectra are compared with simulated data for 7, 8, 9 and 14-nm-thick quantum well heterostructures to gain insight into the physics of the system. 
We find that there is considerable splitting of the $\Delta$-valley conduction band edges, and scattering into four of the six strain-split $\Delta$-valley states is energetically allowed. 
This results in a reduction of the $\Gamma$-valley lifetime compared with bulk unstrained Ge (and with Ge quantum wells grown on SiGe virtual substrates with a smaller Si fraction), and is manifest in the spectra as an increased lifetime broadening component of the excitonic peaks, as well as an increase in the indirect absorption.  We study the excitonic broadening as a function of the applied electric field, allowing separation of the contributions due to lifetime broadening, which is attributed to phonon scattering, and to disorder broadening caused by fluctuations in the dimensions of the quantum wells throughout the structure.

\begin{figure*}
\subfigure[]{
\includegraphics*[width=8cm]{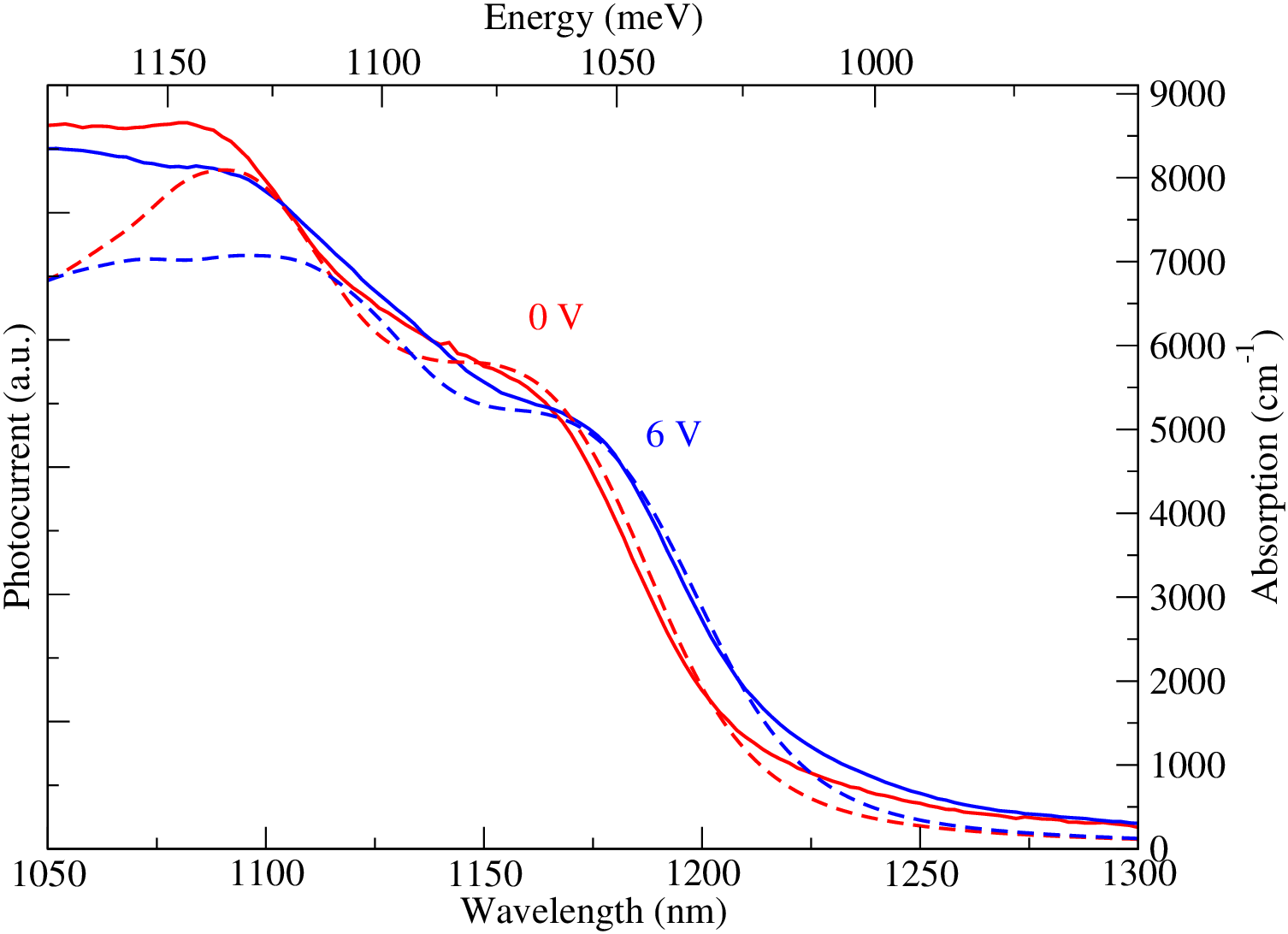}
}
\subfigure[]{
\includegraphics*[width=8cm]{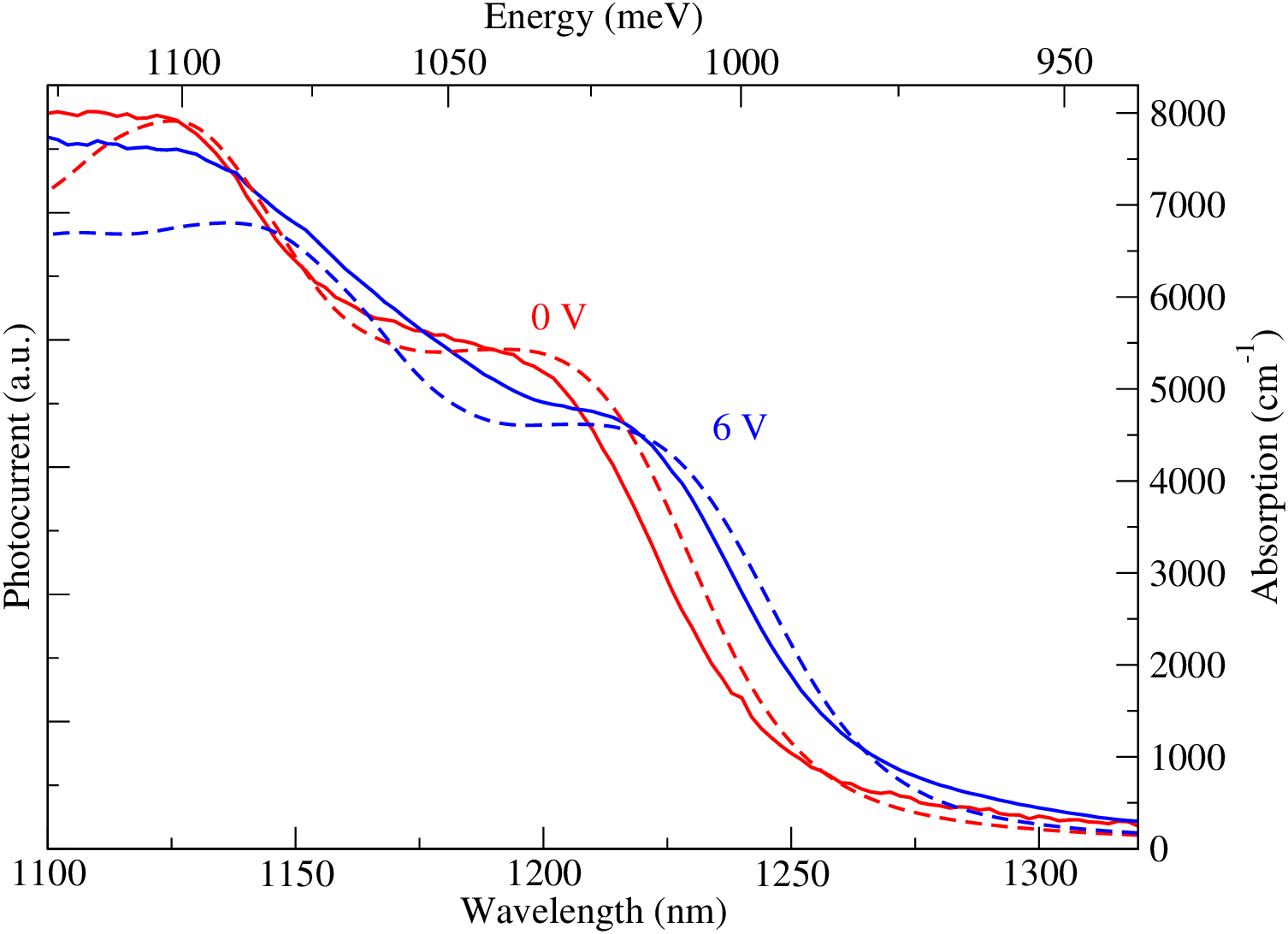}
}
\subfigure[]{
\includegraphics*[width=8cm]{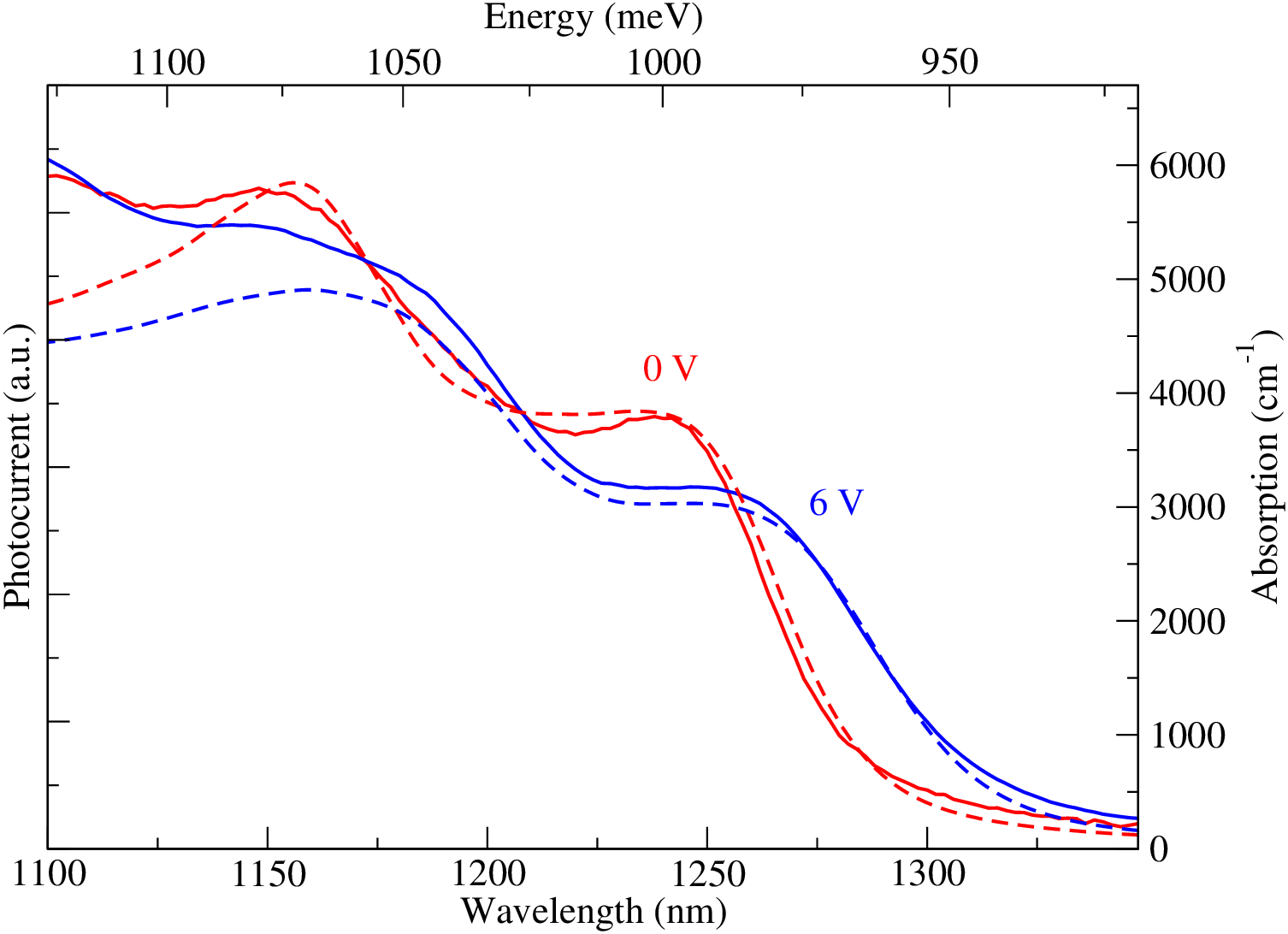}
}
\subfigure[]{
\includegraphics*[width=8cm]{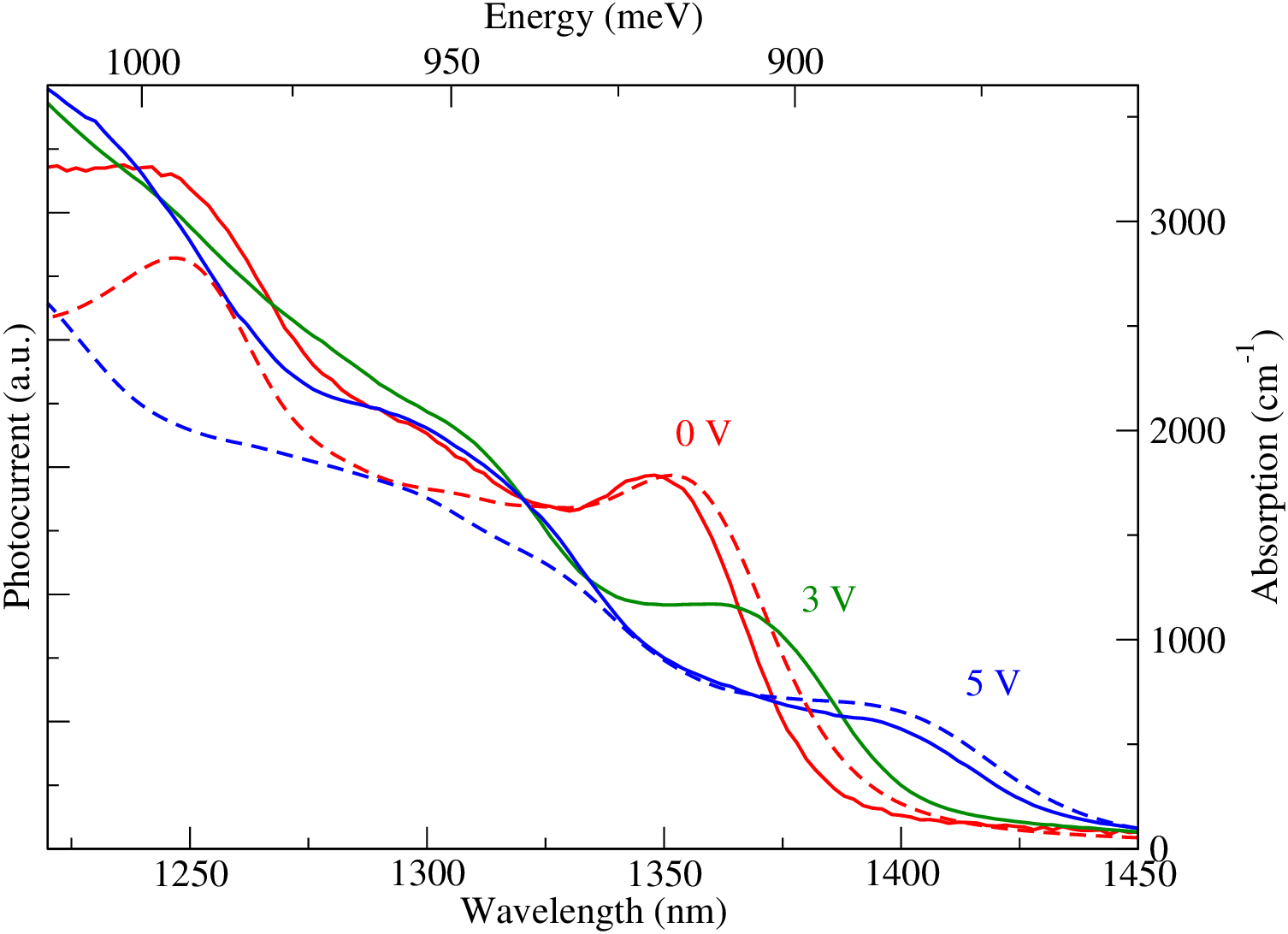}
}
 \caption{The measured photocurrent spectra for the quantum well systems.  (a) 7\,nm wells with 5\,nm barriers, (b) 8\,nm wells with 6\,nm barriers, (c) 9\,nm wells with 7\,nm barriers and (d) 14\,nm wells with 11\,nm barriers.  At zero applied bias, the primary absorption edge at wavelengths of approximately  (a)1160\,nm, (b) 1200\,nm, (c) 1240\,nm, and (d) 1350\,nm corresponds to the hh$_1$-$\Gamma_1$ exciton in the four structures; the features at approximately 1080\,nm, 1125\,nm, 1150\,nm and 1240\,nm correspond to the lh$_1$-$\Gamma_1$ excitons.  Simulated data at zero bias and at the largest measured bias are shown by the dashed lines, and the magnitude of the simulated data is indicated on the right-hand side axis.  Note that the measured data in (c) is the same as that reported in Ref.~\onlinecite{Lever2011}.}
	\label{fig:measured-spectra}
\end{figure*}

\section{Absorption spectra for highly-strained MQW structures}
\label{sec:spectra}
Strain-balanced Ge/Si$_{0.4}$Ge$_{0.6}$ multiple-quantum-well (MQW) heterostructures with ten wells and eleven barriers were grown at 450$^\circ$C using reduced-pressure chemical vapour deposition (RP-CVD) on a relaxed Si$_{0.22}$Ge$_{0.78}$ virtual substrate, which was formed using a reverse linear grading technique.\cite{Shah2008,Liu2011}  The bottom layer of the device was $p$-doped with boron at $1\times10^{19}$\,cm$^{-3}$, on which was grown a 100-nm-thick intrinsic SiGe spacer, the MQW layers, a second 100-nm-thick intrinsic spacer layer, an $n$-type SiGe layer doped with phosphorous at $1\times10^{19}$\,cm$^{-3}$, and finally a 2-nm-thick Si cap layer.  80-$\mu$m-diameter circular mesas were etched to form vertical $p$-$i$-$n$ diodes, and a Ti/Al metal stack was deposited and sintered at 400$^\circ$C for 30~minutes to form electrical contacts.  

A 100-W xenon lamp was used as a light source, and a 3-nm-bandwidth monochromator was used to provide wavelength selectivity.  The signal was optically-chopped  and focussed onto an optical fibre to illuminate the reverse-biased diodes.  Lock-in detection was used to improve the signal-to-noise ratio, allowing the absorption spectra to be inferred from the wavelength-dependent photocurrent measured in the devices.

The photocurrent spectra for the 7, 8, 9 and 14-nm-wide quantum well structures are shown in Figs.~\ref{fig:measured-spectra}(a--d). As expected, there was a systematic blue shift in the absorption edge as the width of the quantum wells is decreased.  The Stark shift at a given applied bias was also larger for the wider quantum wells, as was the reduction in the magnitude of the hh$_1\Gamma_1$ exciton peak that occurs at the absorption edge when a given bias is applied, which results from the larger spatial separation of the electron and hole wavefunctions.  

When the dimensions of the quantum wells were decreased, the amount of excitonic broadening increased, as did the amount of indirect absorption (at a given energy from the direct absorption edge).  This suggests that the increased confinement energy of the $\Gamma$-valley electrons results in a decrease in the $\Gamma$-valley electron lifetime in the quantum well system.  In the following section we describe the energy dependence of the $\Gamma\rightarrow L$ and $\Gamma\rightarrow\Delta$ scattering lifetimes, and go on to relate this to observable features in the absorption spectra, resulting in the simulated spectra that are shown by the dashed curves in Fig.~\ref{fig:measured-spectra}.

\section{$\Gamma$-valley lifetime calculation}
\label{sec:rates}

\begin{figure}
\centering
\includegraphics*[width=8cm]{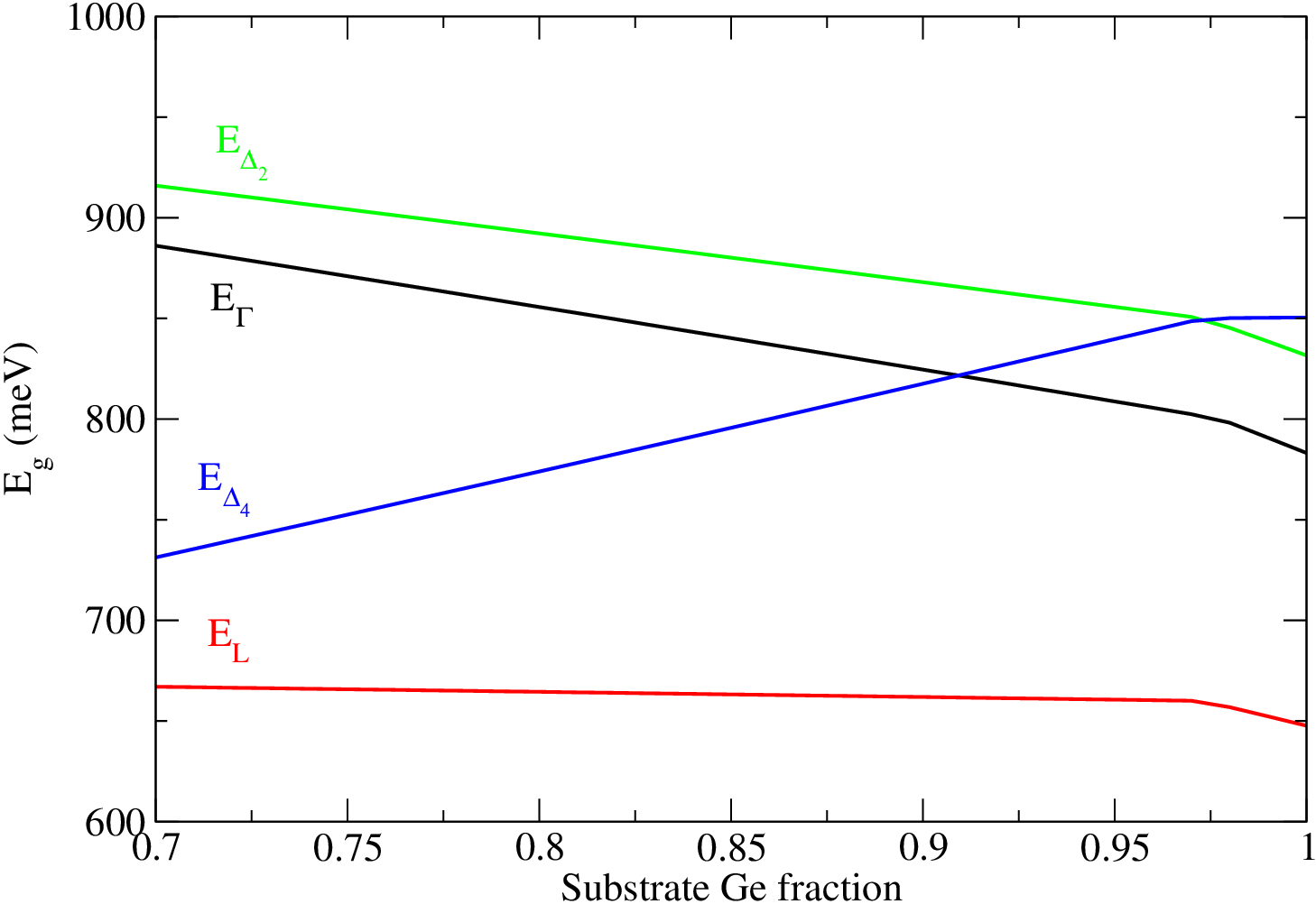}
 \caption{The energy of the conduction band valleys of epitaxial strained Ge relative to the highest energy valence band edge as a function of the virtual substrate composition.  The discontinuity at a Ge fraction of around 97\% arises from the switch between compressive and tensile strain due to the 0.1\% residual tensile strain we have included,\cite{Lever2010} as when there is compressive strain the heavy-hole band edge is highest in energy and when there is tensile strain the light-hole band edge is highest.}
	\label{fig:bandstructure}
\end{figure}


The bandstructure of the strained heterostructure system was calculated using model solid theory with the parameters given in Table~\ref{tab:model-solid-parameters} in the appendix using the method described in Ref.~\onlinecite{Lever2010}.  All calculations were performed assuming room temperature conditions of 300\,K.  
The energy gaps between the highest of the strain-split valence band edges and the different conduction band valleys for epitaxially-grown strained Ge are plotted in Fig.~\ref{fig:bandstructure} as a function of the composition of the virtual substrate composition.  There is a considerable splitting between the in-plane ($\Delta_4$) and the growth axis ($\Delta_2$) $\Delta$-valley states with increasing in-plane compressive strain, as well as an increase in the $\Gamma$-valley band gap.
This means that scattering from the $\Gamma$-valley conduction band states into the $\Delta_4$-valleys is allowed for bulk germanium epitaxially grown on a virtual substrate with a Ge fraction of around 90\% or less.  
For the Ge/Si$_{0.4}$Ge$_{0.6}$ quantum wells on a Si$_{0.22}$Ge$_{0.78}$ virtual substrate, we find a $\Gamma$-valley conduction band offset of 985\,meV, and heavy- and light-hole band offsets of 306\,meV and 161\,meV, respectively.

We used an effective mass model to calculate the confined state electron wavefunctions\cite{Cooper2010} with box boundary conditions to contain the Ge/SiGe finite quantum well structure.  
Intervalley phonon scattering rates from the $\Gamma$-valley into the $L$ and $\Delta$ conduction band valleys were evaluated using a zeroth order deformation potential approach using the longitudinal acoustic and optical phonon scattering parameters given in Ref.~\onlinecite{Tyuterev2011}.  In order to determine the significance of quantum confinement of the electronic states on the scattering rates we compare two approaches: one whereby we treat the initial and final states as subbands, with the variation along the growth-axis described by the confined state wavefunctions of the system, and one where we use the expressions for deformation potential scattering in bulk Ge.

In the intersubband approach, the $\Gamma\rightarrow L$ and $\Gamma\rightarrow \Delta$ scattering rates were determined from 
\begin{eqnarray}
  W_{\Gamma\rightarrow L,\Delta} & = & \frac{2\pi}{\hbar} \frac{\hbar {D_{\Gamma v}}^2 N_{\textrm{val}}}{2\rho\omega_{\Gamma v}} (N_0 + \nicefrac{1}{2} \pm \nicefrac{1}{2})  \nonumber \\
  {} & {} & \times 
  \sum_{j=1}^{N}{  \int_0^{\frac{2\pi}{a}} {\left(\int{ \psi_\Gamma {\psi_j}^* \textrm{e}^{\textrm{i}qz}\textrm{d}z} \right)}^2 \textrm{d}q}     \nonumber \\ 
  {} & {} &  \times  \rho_v(E_k^j\mp\hbar\omega_{\Gamma v}), \label{eqn:conf}
\end{eqnarray}
where the subscript $v$ refers to the destination valley (i.e., either $L$, $\Delta_2$ or $\Delta_4$), so that
$D_{\Gamma \Delta}$ and $D_{\Gamma L}$ are the deformation potentials, $\omega_{\Gamma L}$ and $\omega_{\Gamma \Delta}$ are the phonon frequencies, $N_0$ is the relevant Bose-Einstein phonon occupation number, $\rho$ is the mass density, $q$ is the phonon wavevector, $a$ is the lattice constant, $E_k^j$ is the kinetic energy in the $j^\textrm{th}$ destination subband, $N_{\textrm{val}}$ is the number of equivalent destination valleys, and $\rho_{v}(E_k^j\mp\hbar\omega_{\Gamma v})$ is the density of states of the destination subband, where $\pm$ refers to emission or absorption of a phonon.
The summation is over all energy-conserving destination subbands for a given process and $\psi_\Gamma$ and $\psi_j$ are the initial ($\Gamma$-valley) and final ($L$, $\Delta_2$ or $\Delta_4$-valley) confined-state wavefunctions, respectively.  

In the continuum approach the scattering rates were calculated using the following relation, which describes the $\Gamma$-valley scattering rates in a bulk semiconductor,\cite{Ridley1999}
\begin{eqnarray}
 W_{\Gamma\rightarrow L,\Delta} & = & \frac{2\pi}{\hbar}  
 \frac{\hbar {D_{\Gamma v}}^2 N_\textrm{val}}{2\rho\omega_{\Gamma v}} 
 (N_0 + \nicefrac{1}{2} \pm \nicefrac{1}{2}) \nonumber \\ 
  {} & {} & \times \rho_v(E_k\mp\hbar\omega_{\Gamma v}), \label{eqn:cont}
\end{eqnarray}
where $E_k$ is the final-state kinetic energy of the destination valley for a given scattering process.  Note that here $\rho_v$ is the 3D density of states in the destination valleys.

In both approaches the $\Gamma$-valley inverse scattering lifetime was found by summing the three different rates, i.e.,
\begin{equation}
\nicefrac{1}{\tau_\Gamma} = W_{\Gamma\rightarrow L} + W_{\Gamma\rightarrow \Delta_2} + W_{\Gamma\rightarrow \Delta_4}.
\end{equation}

\begin{figure}
\centering
\includegraphics*[width=8cm]{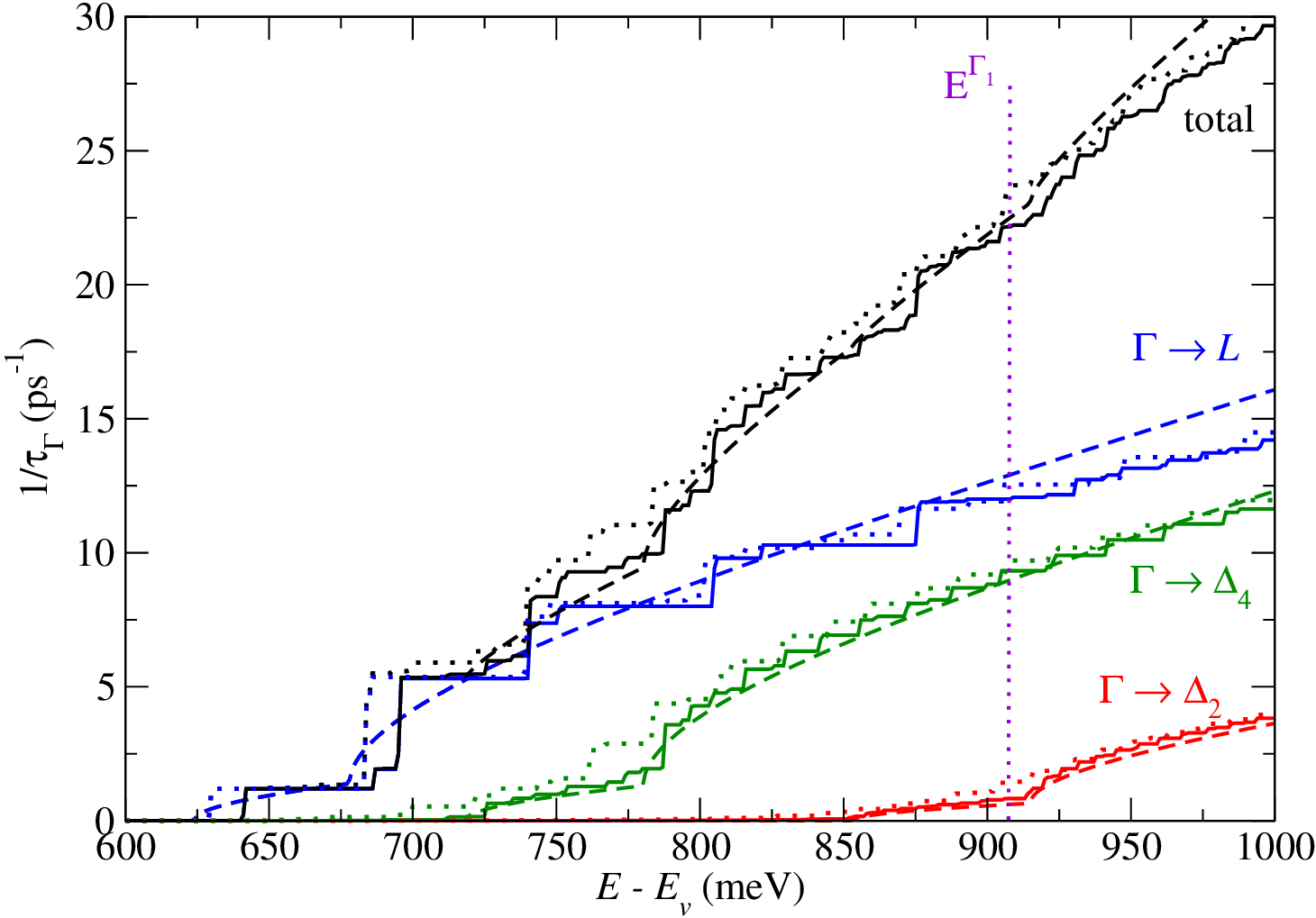}
 \caption{The inverse $\Gamma$-valley scattering lifetime components for intervalley scattering into $L$, $\Delta_2$ and $\Delta_4$ as a function of the energy of the $\Gamma$-valley subband relative to the top of the valence band.  The solid curves show the rates calculated according to the quantum-confined scattering model at zero field and the dashed curves show the rates calculated according to the continuum scattering model.  Also shown in the dotted curves are the scattering rates calculated according to the quantum-confined at 100\,kV/cm.  The system is a 14-nm-thick Ge quantum well with Si$_{0.4}$Ge$_{0.6}$ barriers on a Si$_{0.22}$Ge$_{0.78}$ virtual substrate, and the energy of the $\Gamma_1$ subband is indicated for reference.}
	\label{fig:compare-rates}
\end{figure}

Figure~\ref{fig:compare-rates} shows the scattering rates obtained using both models. 
The effects of quantum confinement are reflected in the scattering rates, with the step-like features that result from the quantum-confined density of states of the destination subbands.  At zero field the onset of phonon absorption into the first $L$-valley subband is seen at 641\,meV and phonon emission scattering into this subband at 696\,meV.  With an applied field these features occur energies approximately 10\,meV smaller, reflecting the Stark shift of the $L$-valley confined-state wavefunctions.   The small step at 687\,meV results from the onset of phonon absorption into the second $L$-valley subband. At energies larger than $\sim750$\,meV the quantum-confined nature of the scattering rates is no longer apparent, as multiple destination subbands are energetically allowed for the scattering processes.
There is good agreement between the two models in terms of the overall magnitude of the rates (less than 10\% deviation across a range of quantum well dimensions and substrate alloy fractions); the principal difference results from the density of final states, whilst the details of the wavefunctions themselves, as well as the overlaps between them, appear to be unimportant.  
Therefore, due to the relative simplicity of implementing equation \ref{eqn:cont}, we choose to use the continuum method for the subsequent investigation of the $\Gamma$-valley scattering lifetime.

We model indirect absorption according to the method described in Ref.~\onlinecite{Ridley1999}, so that we have 
\begin{eqnarray}
\alpha_{\textrm{in}}(\omega) &=& 
C_0 S_{\textrm{2D}} \frac{2}{L} \frac{1}{4\pi} \frac{\hbar}{2\pi} 
 \iint {\left|  {\Bra{\psi_\Gamma} \textrm{\textbf{e}} \cdot \textrm{\textbf{p}}\Ket{\psi_h}}\right|}^2 
\nonumber \\ 
& & \times  \frac{1}{(E_{0} - \hbar\omega)^2 }  \frac{1}{\tau_\Gamma}
\textrm{d}k_x \textrm{d}k_y, 
\label{eqn:alpha_in}
\end{eqnarray}
where $S_{\textrm{2D}}$ is the Coulomb enhancement factor, $L$ is the periodicity of the MQW system (i.e., the well width plus the barrier width).  The term ${\left|  {\Bra{\psi_\Gamma} \textrm{\textbf{e}} \cdot \textrm{\textbf{p}}\Ket{\psi_h}}\right|}$ is the momentum matrix element between the $\Gamma$-valley electron and hole wavefunctions, which was evaluated using the formalism in Ref.~\onlinecite{Chang1995}; at the Brillouin zone centre, the reduces to
\begin{equation}
{\left|  {\Bra{\psi_\Gamma} \textrm{\textbf{e}} \cdot \textrm{\textbf{p}}\Ket{\psi_h}}\right|}^2 	
	= E_p m_0 \left(\int {{\psi_\Gamma} {\psi_h}^*}\textrm{d}z\right)^2,
\label{eqn:matrix-element}
\end{equation}
where $E_p=26.3$\,eV is the optical transition matrix element for germanium\cite{Ridene2001} and $m_0$ is the electron rest mass.  (Away from the zone centre the expression becomes more complex as the hole wavefunctions are given by a linear combination of light-hole, heavy-hole, and spin-orbit split-off basis states.)  
The terms $k_x$ and $k_y$ are the wavevectors in the growth plane, the factor of two accounts for spin degeneracy, and $C_0$ is given by
\begin{equation}
	C_0 = \frac{\pi e^2}{N_r c \epsilon_0 {m_0}^2 \omega},
\end{equation}
where $\omega$ is the angular frequency of the light and $N_r$ is the real part of the refractive index of the material. 
The term $E-\hbar\omega$ is the shortfall between the photon energy and the energy of the intermediate $\Gamma$-valley state, and $\tau_\Gamma$ is the lifetime of the intermediate $\Gamma$-valley state at that energy,
i.e., the energy corresponding to the absorption of light with a photon energy smaller than the direct absorption edge.
It is for this reason that we determine the intervalley scattering rates as a function of the energy of the $\Gamma$-valley state relative to the top of the valence band for both the confined and continuum methods, i.e., for a range of energies including those smaller than the $\Gamma$-valley band gap.

The excitonic contribution to the spectra from each pair of electron and hole subbands was modelled according to
\begin{eqnarray}
 \alpha_{\textrm{ex}} & = & \frac{2C_0}{L} 
{\left|  {\Bra{\psi_\Gamma} \textrm{\textbf{e}} \cdot \textrm{\textbf{p}}\Ket{\psi_h}}\right|}^2
\frac{2}{\pi\lambda^2} \nonumber \\
& & \times V(E_{i,j}-E_b, \hbar\omega, \gamma_\tau, \gamma_\sigma)
\end{eqnarray}
where $E_{i,j}$ is the energy difference between a given pair of subband minima, $E_b$ is the exciton binding energy, $\lambda$ is the exciton Bohr radius, which was determined using a variational energy minimisation approach\cite{Fox1991,Susa1993} and $V()$ is the Voigt profile, where $\gamma_\tau$ and $\gamma_\sigma$ are the lifetime and disorder broadening components.  


\section{Discussion}
\label{sec:discussion}
Simulated absorption spectra for the four structures are shown by the dashed lines in Figs.~\ref{fig:measured-spectra}(a--d) at zero applied bias and also at the maximum of the applied biases, $V_\textrm{max}$.  
The simulated datasets were calculated based on the method described in Ref.~\onlinecite{Lever2010} and the parameters listed in the appendix (Table~\ref{tab:model-solid-parameters}), together with the methodology described above, which were used to determine the lifetime broadening components of the exciton peaks as well as the indirect absorption. 
We find very good agreement in the wavelengths of the absorption edge caused by the hh$_1\Gamma_1$ exciton, and the small shifts from the experimentally-measured absorption edge are consistent with a deviation of around 1\,\AA{} from the target well width. 

We account for interdiffusion of the Ge and SiGe layers by describing the alloy composition across the Ge/SiGe interface using the Gauss error function.\cite{Li1996,Lever2010}  There are a number of values available in the literature for most of the parameters listed in Table~\ref{tab:model-solid-parameters}, and it is possible to reproduce the measured absorption edge wavelength for the structures reported here using a number of different sets of data.  However, in order to reproduce the light-hole/heavy-hole splitting and the absorption edge for our structures as well as other structures reported previously (including those in Refs.~\onlinecite{Kuo2005,Kuo2006,Schaevitz2008,Lange2009}, where both Si$_{0.1}$Ge$_{0.9}$ and Si$_{0.05}$Ge$_{0.95}$ virtual substrates have been used), the range of acceptable parameters is considerably reduced.  We found that those parameters listed in Table~\ref{tab:model-solid-parameters}, together with an interdiffusion length of 2\,nm for the structures reported here, gave agreement with the absorption edge of the available experimental data.\cite{Lever2011}

The magnitude of the measured sub-band-gap absorption shown in Figs.~\ref{fig:measured-spectra}(a--d) increased as the width of the quantum wells decreased.  This trend is also present in the simulated indirect absorption in the four different structures, where the increase in the relative magnitude of the indirect absorption results from the increase in $W_{\Gamma\rightarrow L,\Delta}$. This happens because the $\Gamma$-valley ground subband energy is larger in the smaller quantum wells, and hence there will be a larger density of final states in the $L$- and $\Delta$-valleys available for electrons to scatter into. 
The magnitude of the measured sub-band-gap absorption was larger than in the simulated data, especially in the smaller quantum wells.  This may result from the effect of threading dislocations, which have been shown to cause sub-band-gap absorption in highly-strained heteroepitaxial systems with large defect densities.\cite{Peiner2002} 

There are two components of the excitonic broadening: one arising from structural disorder or inhomogeneities, and one arising from the finite lifetime of the $\Gamma$-valley states.  When comparing to the simulated data it is not straightforward to separate the two, because the Voigt profile used to model the excitonic broadening is a convolution of a Gaussian with a Lorentzian, and so the sum of linewidth broadening components does not equal the resulting linewidth. 
Our approach to reproducing the measured broadening of the absorption edge was to determine the lifetime broadening contribution to the exciton linewidths using the calculated $\Gamma$-valley scattering lifetime, and then to adjust the disorder broadening component so that the resulting spectrum has the same lineshape as the measured spectra, giving the simulated datasets shown by the dashed lines in  Fig.~\ref{fig:measured-spectra}.

\begin{table}
\caption{The calculated $\Gamma$-valley lifetime and associated HWHM of the hh$_1$-$\Gamma_1$ exciton, as well as the disorder broadening components required to reproduce the measured data. }
\label{tab:HWHM}
\begin{center}
	\begin{tabular}{p{1.5cm}p{1.6cm}p{1.6cm}p{1.6cm} p{1.6cm}}
\hline
\hline
\footnotesize{Well width (nm)} & 
\footnotesize{Calculated lifetime, $\tau_\Gamma$ (fs)} & 
\footnotesize{Calculated lifetime broadening, $\gamma_\tau$ (meV)} & 
\footnotesize{Inhomo\-gen\-eous broadening, $\gamma_\sigma$,  at 0\,V (meV)}  & 
\footnotesize{Inhomo\-gen\-eous broadening, $\gamma_\sigma$, at $V_\textrm{max}$ (meV)}\\
\hline
7 &   29.6 &  11.1 &  17    & 18 \\
8 &   33.6 &  9.8  &  15    & 16 \\
9 &   36.2 &  9.1  &  11    & 14 \\
14 &  47.0 &  7.0  &   7    & 11 \\
\hline
\hline
\end{tabular}
\end{center}
\end{table}

\begin{figure}
\includegraphics*[width=8cm]{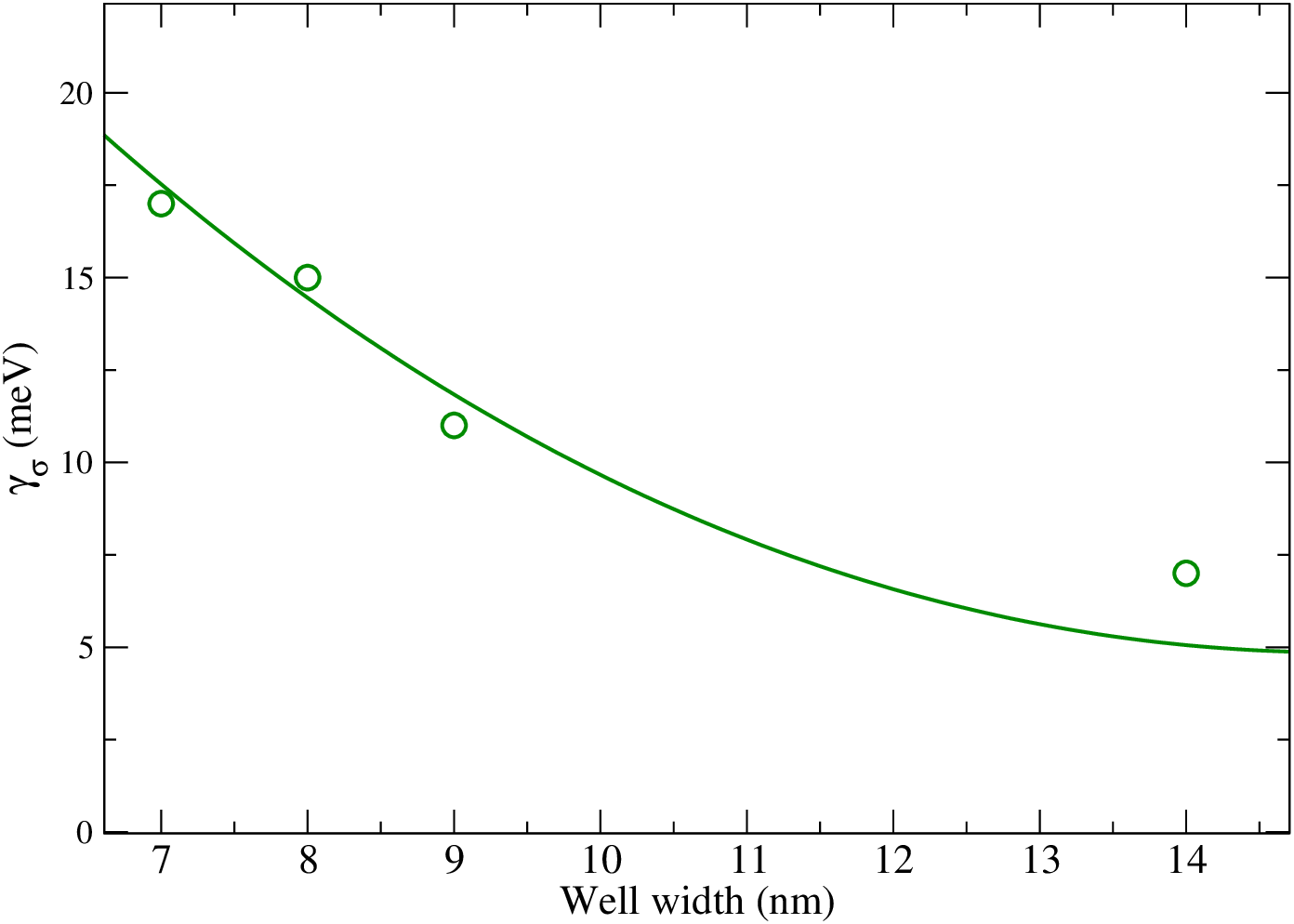}
 \caption{The disorder broadening HWHM as a function of the well width.  The data points are the HWHM disorder broadening parameters at $V=0$ from Table~\ref{tab:HWHM}, and the curve was generated using equation~\ref{eqn:zimmermann} with $h\xi=2.4$\,nm$^2$.}
	\label{fig:sigma-vs-L}
\end{figure}

The broadening of the absorption edge in the four measured samples is characterised by the half-width at half-maximum (HWHM) and scattering lifetime data listed in Table ~\ref{tab:HWHM}.  The lifetime broadening reflects the $\Gamma$-valley scattering lifetime and, as expected, increased with increasing confinement energy.  
The disorder broadening component increased as the width of the quantum wells was reduced.  Additionally, for the larger quantum wells, the disorder broadening component also increased at the larger applied biases.  Disorder broadening resulting from fluctuations in the thickness of the quantum wells throughout the structure may be described according to\cite{Zimmermann1997}
\begin{equation}
\frac{{\gamma_\sigma}^2}{2\textrm{ln}(2)} = \frac{2 h^2\xi^2}{\lambda^2} 
\left[ 
\left( E_e' \beta_e \right)^2 + 
8 E_e' E_h' + 
\left( E_h' \beta_h \right)^2
\right]
\label{eqn:zimmermann}
\end{equation}
where $E_e'$ and $E_h'$ are the differential of the $\Gamma$-valley electron and heavy-hole confinement energies with respect to the well width, and $\beta_e$ and $\beta_h$ relate the electron and hole in-plane effective masses as follows:
\begin{equation}
\beta_e = \frac{m_e + m_h}{m_h}, \qquad \beta_h = \frac{m_h + m_e}{m_e}.
\end{equation}
The terms $h$ and $\xi$ are the height and correlation length of the interface disorder. 
Figure \ref{fig:sigma-vs-L} shows the disorder broadening parameters for $V=0$ listed in Table~\ref{tab:HWHM}, together with HWHM extracted from equation \ref{eqn:zimmermann} using the simulated values for $E_e'$, $E_h'$ and $\lambda$.  
It can be seen that the disorder broadening extracted from the experimental data follows the expected form given by equation~\ref{eqn:zimmermann} and, furthermore, that the fit obtained with $h\xi=2.4$\,nm$^2$ provides a measure of the interface roughness in these samples.  

\begin{figure}
\includegraphics*[width=8cm]{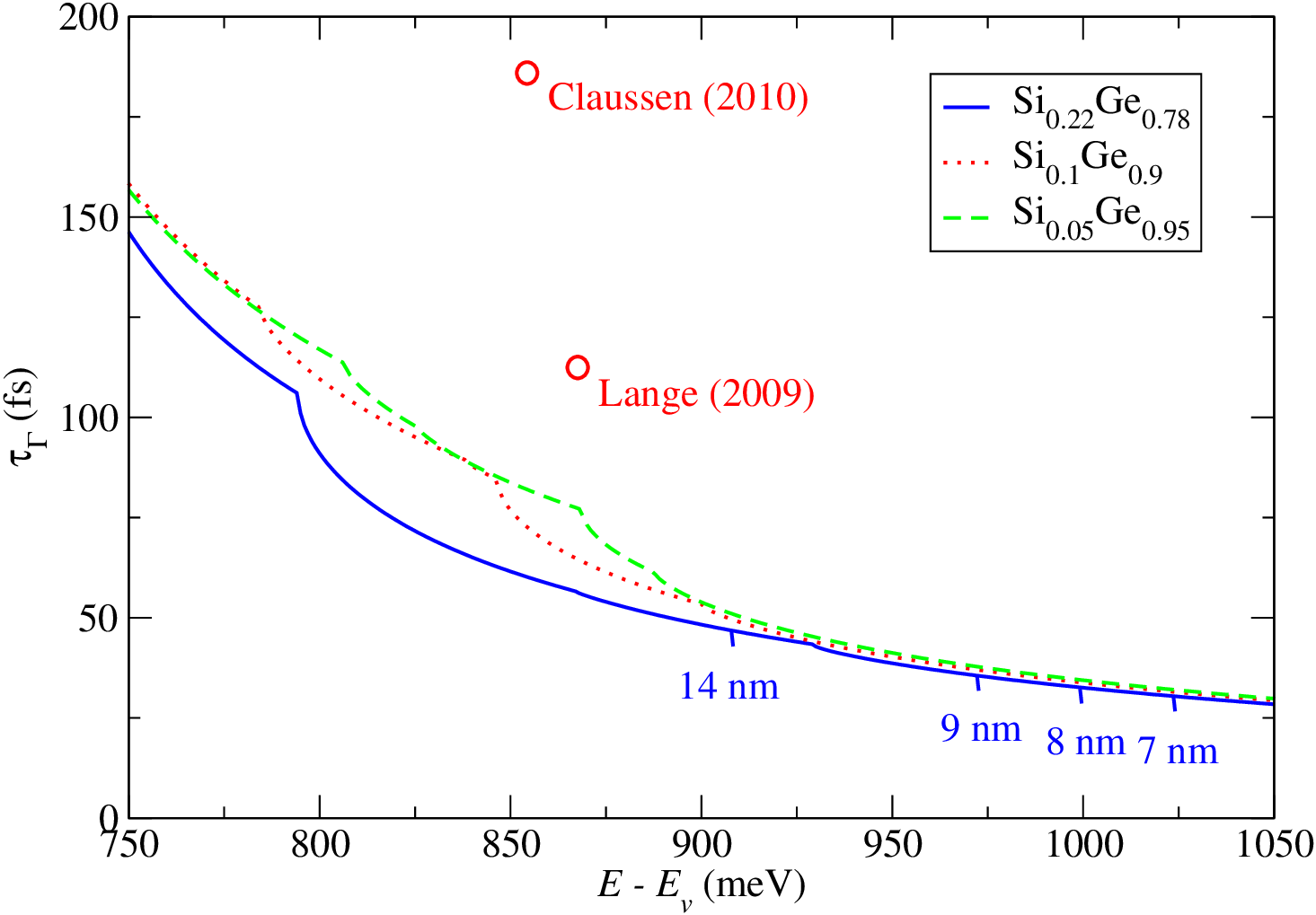}
 \caption{
The $\Gamma$-valley scattering lifetime is shown in for each of the substrate compositions.  The two available measured lifetimes for Ge/SiGe quantum wells on a  Si$_{0.1}$Ge$_{0.9}$ virtual substrate are shown for comparison, based on data from Lange \emph{et al.}\cite{Lange2009} and Claussen \emph{et al.}\cite{Claussen2010}  Additionally, the quantum well systems reported here are included on the curve for the Si$_{0.22}$Ge$_{0.78}$ virtual substrate.  Note that these values are the strained band gap plus the electron confinement energy; to compare the results of these calculations with the photon energy, we must add on the confinement energy of the hole states.}
	\label{fig:rates}
\end{figure}


In order to gain more insight into how strain affects the $\Gamma$-valley lifetime in Ge/SiGe quantum well systems, the intervalley scattering lifetimes are shown as a function of energy in Fig.~\ref{fig:rates} for Ge fractions in the virtual substrate of 78\%, 90\%, and 95\% (i.e., those substrate compositions where the QCSE has been demonstrated).  As expected, for a given substrate composition, a larger $\Gamma$-valley energy (corresponding to a smaller quantum well) results in more rapid scattering.  
At a given energy the $\Gamma$-valley lifetime is longest with the Si$_{0.05}$Ge$_{0.95}$ virtual substrate where there is less compressive strain.  Furthermore, the effect is most pronounced in the region around 800--900\,meV, which is the range where the QCSE has been demonstrated in Ge/SiGe MQW systems.  This results mainly from the large strain-induced splitting between the $\Delta_2$ and $\Delta_4$ conduction band valleys, which affects the energy at which $\Gamma\rightarrow \Delta$ scattering becomes allowed.  
Moreover, because of the hydrostatic strain shift of the $\Gamma$-valley band gap, for a given confinement energy the value of $E-E_v$ will be smaller for a virtual substrate with a larger Ge fraction.  These two considerations imply that the $\Gamma$-valley lifetime will be maximised by choosing a larger Ge fraction in the virtual substrate, which will result in structures with less indirect absorption and less lifetime broadening of the excitonic peaks.  
The two available measured values for the $\Gamma$-valley lifetime in these systems are indicated in  Fig.~\ref{fig:rates}.  The calculations described here give rates for $W_{\Gamma\rightarrow L,\Delta}$ that are almost twice as fast as have been reported experimentally.  
However, it should be pointed out that these experimental systems are limited in the temporal resolution owing to the finite duration of the optical pump pulses, and furthermore rely on optical bleaching (saturation), which has implications for final state blocking effects in the scattering processes.  Consequently, it is possible that these measurements may over-estimate the $\Gamma$-valley lifetime, and so it is not entirely surprising that the calculated lifetimes reported here are shorter.

\section{Conclusion}
We have measured absorption spectra and observed the QCSE in Ge/SiGe MQW structures grown on Si$_{0.22}$Ge$_{0.78}$ virtual substrates, where the dimensions of the quantum wells were systematically varied.  The large compressive strain in Ge quantum wells results in a significant blue shift of the absorption edge compared with structures that were grown on virtual substrates with larger Ge fractions.  Additionally, a broadening of the exciton peaks is observed, as well as an increase in the magnitude of the indirect absorption.

We calculated the intervalley phonon scattering rates in Ge/SiGe MQW systems in order to determine the $\Gamma$-valley lifetime.  Two methods were compared: one accounting for the quantum confinement of the electron wavefunctions and one assuming bulk-like wavefunctions.  Good agreement was found between the two approaches, indicating that the details of the quantum confined wavefunctions are unimportant for scattering, and that the rates are primarily determined by the density of final states.  

Indirect absorption in unstrained bulk Ge results from $\Gamma\rightarrow L$ intervalley phonon scattering, because scattering into the $\Delta$-valleys is not energetically allowed.  However, in strain-balanced Ge/SiGe MQW stacks, the combined effects of the in-plane compressive strain and the confinement energy of the $\Gamma$-valley electrons means that $\Gamma\rightarrow \Delta$ scattering can occur, resulting in a decrease in the $\Gamma$-valley electron lifetime compared with bulk unstrained Ge.
We find that $\Gamma\rightarrow\Delta$ scattering is an important contribution to the $\Gamma$-valley lifetime, and the calculated lifetimes in these highly-strained Ge/SiGe quantum wells were found to be less than 50\,fs. 

Although technological issues resulting in threading dislocations and interface roughness appear likely to affect the exciton linewidths, the results of our calculations imply that reducing the compressive strain in the Ge quantum wells by choosing a virtual substrate composition with a large Ge fraction will increase the $\Gamma$-valley lifetime because of the effects of intervalley phonon scattering.  This will reduce the indirect absorption and lifetime broadening of the excitonic peaks, leading to an improvement in the performance that can be achieved in an EAM based on Ge/SiGe quantum wells.  While the performance at 1.3\,$\mu$m can be improved by reducing the sources of inhomogeneous broadening via improved growth, this is also the case for structures operating at longer wavelengths,  and our work indicates that the wavelength-dependent performance of Ge/SiGe modulators will remain in place.

\appendix

\section{Material parameters}
Table \ref{tab:model-solid-parameters} lists the material parameters used to generate the simulated data:  $\gamma_1$, $\gamma_2$ and $\gamma_3$, are the Luttinger parameters used to calculate the valence band structure; $\Delta_\textrm{SO}$ is the spin-orbit splitting; $a_\textrm{latt}$ is the lattice constant; $m^\Gamma$ is the $\Gamma$-valley electron effective mass, $m_l^L$ and $m_l^{\Delta}$ is the longitudinal effective masses in the $L$- and $\Delta$-valleys, respectively, and $m_t^L$ and $m_t^\Delta$ are the transverse effective masses; $E_\Gamma$, $E_L$ and $E_\Delta$ are the unstrained band gaps; $a^\Gamma$, $a^L$ and $a^\Delta$ are the hydrostatic band gap deformation potentials, and $b$ and $b^\Delta$ are the shear band gap deformation potentials (owing to the symmetry of the system there is no splitting of the $L$-valleys in response to biaxial in-plane strain); $D_{\Gamma \Delta}$ and $D_{\Gamma L}$ are the intervalley phonon scattering deformation potentials and $\omega_{\Gamma L}$ and $\omega_{\Gamma \Delta}$ are the phonon angular frequencies.  Where applicable, linear interpolation was used to determine the material parameters of the SiGe alloy layers, except for the lattice constant, where a bowing parameter of 0.026\,nm was used.\cite{DeSalvador2000}   A nonparabolicity parameter of 0.3\,eV$^{-1}$ was used to determine the final density of states in the scattering rate calculations.\cite{Jacaboni1981}

\begin{table}
\caption{The material parameters used in the calculations. } 
\label{tab:model-solid-parameters}
\begin{center}
	\begin{tabular}{ p{3cm} r @{.} l c r @{.} l }
\hline
\hline
Quantity  & \multicolumn{2}{c}{Si} & $\qquad$ $\qquad$  &  \multicolumn{2}{c}{Ge} \\
\hline

 $\gamma_1$ &  4&22$^a$ & & 13&4$^a$ \\
 $\gamma_2$ &  0&39$^a$ & & 4&25$^a$ \\
 $\gamma_3$ &  1&44$^a$ & & 5&69$^a$ \\
\hline
 $\Delta_\textrm{SO}$ (eV)   & 0&044$^a$ & & 0&296$^a$ \\
\hline
 $a_\textrm{latt}$ (nm) & 0&5431$^a$ & & 0&5657$^a$ \\
\hline
 $m^\Gamma$   &  0&156$^b$ & & 0&042$^c$ \\
 $m_l^L$        & 1&7$^d$ & & 1&64$^e$ \\
 $m_l^{\Delta}$   &  0&91$^f$ & & 1&353$^g$ \\
 $m_t^L$         & 0&12$^d$  & & 0&08$^h$ \\
 $m_t^{\Delta}$    & 0&19$^h$ & & 0&288$^g$ \\
\hline
 $E_\Gamma$ (eV) &   4&2$^i$     & &  0&8$^i$  \\
 $E_L$ (eV)      &   2&0$^i$ & & 0&66$^i$ \\
 $E_\Delta$ (eV) &   1&2$^i$  & & 0&85$^i$ \\
\hline  
 $a^\Gamma$ (eV) &    -11&39$^j$        & & -8&97$^k$ \\
 $a^L$ (eV)      &    -3&12$^l$   & & -2&78$^l$ \\
 $a^\Delta$ (eV) &    1&72$^l$   & & 1&31$^l$ \\
 $b$ (eV)  &  -2&1$^m$ & & -2&86$^n$ \\
 $b^\Delta$ (eV)  &  9&61$^l$ & & 9&42$^l$ \\
\hline
 $D_{\Gamma \Delta}$ (eV/cm) & \multicolumn{2}{c}{-} & & 2 & ${5\times10^8}^o$\\ 
 $D_{\Gamma L}$ (eV/cm) & \multicolumn{2}{c}{-} & & 4 & ${0\times10^8}^o$\\ 
 $\omega_{\Gamma L}$ (THz) & \multicolumn{2}{c}{-} & & 6 & $54\times2\pi^o$ \\
 $\omega_{\Gamma \Delta}$ (THz) & \multicolumn{2}{c}{-} & & 7 & $61\times2\pi^o$ \\
\hline
\hline
$^a$ Data from Ref.~\onlinecite{Kahan1994}  & \multicolumn{2}{c}{\,} & \multicolumn{2}{c}{\,} & \\
$^b$ Data from Ref.~\onlinecite{Fischetti1991}  & \multicolumn{2}{c}{\,}  & \multicolumn{2}{c}{\,} & \\
$^c$ Data from Ref.~\onlinecite{Cardona1966}  & \multicolumn{2}{c}{\,}  & \multicolumn{2}{c}{\,} & \\
$^d$ Data from Ref.~\onlinecite{Rieger1993}  & \multicolumn{2}{c}{\,}  & \multicolumn{2}{c}{\,} & \\
$^e$ Data from Ref.~\onlinecite{Dexter1956}  & \multicolumn{2}{c}{\,}  & \multicolumn{2}{c}{\,} & \\
$^f$ Data from Ref.~\onlinecite{Hensel1965}  & \multicolumn{2}{c}{\,}  & \multicolumn{2}{c}{\,} & \\
$^g$ Data from Ref.~\onlinecite{Jacaboni1981}  & \multicolumn{2}{c}{\,}  & \multicolumn{2}{c}{\,} & \\
$^h$ Data from Ref.~\onlinecite{Dresselhaus1955}  & \multicolumn{2}{c}{\,}  & \multicolumn{2}{c}{\,} & \\
$^i$ Data from Ref.~\onlinecite{Levinshtein1996}  & \multicolumn{2}{c}{\,}  & \multicolumn{2}{c}{\,} & \\
$^j$ Data from Ref.~\onlinecite{Wei1999}  & \multicolumn{2}{c}{\,}  & \multicolumn{2}{c}{\,} & \\
$^k$ Data from Ref.~\onlinecite{Liu2004}  & \multicolumn{2}{c}{\,}  & \multicolumn{2}{c}{\,} & \\
$^l$ Data from Ref.~\onlinecite{Van-de-Walle1986}  & \multicolumn{2}{c}{\,}  & \multicolumn{2}{c}{\,} & \\
$^m$ Data from Ref.~\onlinecite{Laude1971}  & \multicolumn{2}{c}{\,}  & \multicolumn{2}{c}{\,} & \\
$^n$ Data from Ref.~\onlinecite{Chandrasekhar1977}  & \multicolumn{2}{c}{\,}  & \multicolumn{2}{c}{\,} & \\
$^o$ Data from Ref.~\onlinecite{Tyuterev2011} & \multicolumn{2}{c}{\,}  & \multicolumn{2}{c}{\,} & 
	\end{tabular}
\end{center}
\end{table}


\end{document}